% ...
\documentclass[prl,unsortedaddress,superscriptaddress,showpacs,twocolumn,floatfix]{revtex4}

\usepackage[dvips]{pstcol}
\usepackage{graphicx}
\begin{document}
\title{Paul trapping of radioactive $^6$He$^+$ ions and direct observation
of their $\beta$-decay}

\affiliation{LPC-Caen, ENSICAEN, Universit\'e de Caen, CNRS/IN2P3-ENSI, Caen, France}
\affiliation{GANIL, CEA/DSM-CNRS/IN2P3, Caen, France}

\author{X.~Fl\'echard}
\affiliation{LPC-Caen, ENSICAEN, Universit\'e de Caen, CNRS/IN2P3-ENSI, Caen, France}
\author{E.~Li\'enard}
\affiliation{LPC-Caen, ENSICAEN, Universit\'e de Caen, CNRS/IN2P3-ENSI, Caen, France}
\author{A.~M\'ery}
\altaffiliation[Present address: ]{SUBATECH, Ecole des Mines de Nantes,
44307 Nantes, France.}
\affiliation{LPC-Caen, ENSICAEN, Universit\'e de Caen, CNRS/IN2P3-ENSI, Caen, France}
\author{D.~Rodr\'{i}guez}
\altaffiliation[Present address: ]{Dep. de F\'{i}sica Aplicada, 
Univ. de Huelva, 21071 Huelva, Spain.}
\affiliation{LPC-Caen, ENSICAEN, Universit\'e de Caen, CNRS/IN2P3-ENSI, Caen, France}
\author{G.~Ban}
\affiliation{LPC-Caen, ENSICAEN, Universit\'e de Caen, CNRS/IN2P3-ENSI, Caen, France}
\author{D.~Durand}
\affiliation{LPC-Caen, ENSICAEN, Universit\'e de Caen, CNRS/IN2P3-ENSI, Caen, France}
\author{F.~Duval}
\affiliation{LPC-Caen, ENSICAEN, Universit\'e de Caen, CNRS/IN2P3-ENSI, Caen, France}
\author{M.~Herbane}
\altaffiliation[Present address: ]{Dep. of Physics, College of Science,
King Khalid Univ., Abha-Saudi Arabia.}
\affiliation{LPC-Caen, ENSICAEN, Universit\'e de Caen, CNRS/IN2P3-ENSI, Caen, France}
\author{M.~Labalme}
\affiliation{LPC-Caen, ENSICAEN, Universit\'e de Caen, CNRS/IN2P3-ENSI, Caen, France}
\author{F.~Mauger}
\affiliation{LPC-Caen, ENSICAEN, Universit\'e de Caen, CNRS/IN2P3-ENSI, Caen, France}
\author{O.~Naviliat-Cuncic}
\affiliation{LPC-Caen, ENSICAEN, Universit\'e de Caen, CNRS/IN2P3-ENSI, Caen, France}
\author{J.C.~Thomas}
\affiliation{GANIL, CEA/DSM-CNRS/IN2P3, Caen, France}
\author{Ph.~Velten}
\affiliation{LPC-Caen, ENSICAEN, Universit\'e de Caen, CNRS/IN2P3-ENSI, Caen, France}

\date{\today}

\begin{abstract}
We demonstrate that abundant quantities of
short-lived $\beta$ unstable ions can be trapped in a novel
transparent Paul trap and that their decay products can directly be
detected in coincidence. Low energy $^6$He$^+$ 
(807~ms half-life) ions were extracted from the SPIRAL source
at GANIL, then decelerated, cooled and bunched by means of the buffer
gas cooling
technique. More than $10^8$ ions have been stored over a measuring period
of six days and about $10^5$ decay
coincidences between the beta particles and the $^6$Li$^{++}$ recoiling ions 
have been recorded. The technique can be extended to other 
short-lived species, opening new possibilities for trap assisted decay
experiments.
\end{abstract}


\maketitle
%
Atom and ion traps have found a wide range of applications in nuclear
physics for the confinement of radioactive species \cite{Sprouse97,Bollen02}.
In particular,
the continuous improvements in magneto-optical trapping efficiencies achieved
since more than ten years
\cite{Lu94,Gwinner94,Behr97,Guckert98} resulted in a number
of precision measurements for the study of
fundamental interactions \cite{Crane01,Scielzo04,Gorelov05} as well as
for the determination of nuclear static properties \cite{Wang04,Mueller07}.
The environment
offered by traps in beta decay measurements is ideal to reduce instrumental
effects, like the
electron scattering in matter, or to enable the direct detection of
recoiling
ions. Such conditions led to measurements of angular correlation coefficients
in beta decay with unprecedented
precision \cite{Scielzo04,Gorelov05} motivated by the search for exotic interactions
as signatures of physics beyond the standard electroweak model
\cite{Severijns06}. 

Although the principles for ion trapping \cite{Brown86,Paul90} were established
well before those for the magneto-optical confinement \cite{AtomTraps}, the
consideration of ion traps for 
beta decay experiments with radioactive species is more
recent.
Technically, magneto-optical traps are often limited to alkali elements,
for which suitable lasers can be found. They enable to produce samples
of smaller size and 
with atoms at lower energies than ion traps.
More elaborated transition schemes have recently been applied
to radioactive He atoms \cite{Wang04,Mueller07}. However,
the efficiencies achieved so far with noble gas atoms in MOTs 
are too small for practicable precision measurements of beta decay
correlations.

The standard geometry of a 3D-Paul
trap \cite{Paul90}, in which the hyperbolic electrodes are made of massive
materials,
is not well suited for the detection of decay products following beta decay.
Ion confinement of radioactive species
requires also the beam preparation for
an efficient trapping \cite{Lunney99} and such techniques posed
new challenges when applied to light mass species.

In this paper we demonstrate that significant quantities of
radioactive $^6$He$^+$ ions can be confined in a novel Paul trap and that their
decay products can directly be detected, enabling thereby new trap
assisted decay experiments.
The results presented here were motivated by a new measurement of the
$\beta\bar{\nu}$ angular correlation coefficient
in the Gamow-Teller decay of $^6$He
to search for possible exotic interactions in nuclear beta decay.
For this purpose, beta particles and recoiling ions are detected in coincidence
to deduce the time of flight spectrum of ions relative to the beta particles.

The experiment has been carried out at the new low energy beam line
LIRAT \cite{Varenne05} of the SPIRAL facility at GANIL, 
firstly commissioned with radioactive beams in 2005. 
The $^6$He$^+$ ions were produced from a primary $^{13}$C beam at 75~MeV/A
impinging
on a graphite target coupled to an ECR source. The source is located on a
high voltage platform at 10~kV, what determines the kinetic
energy of the extracted ions.
The beam is then mass separated by a dipole magnet
having a resolving power $M/\Delta M \approx 250$.
After separation in $m/q$, the beam is
composed of $^6$He$^+$ and of stable $^{12}$C$^{2+}$ ions.
By adjusting the setting of the first magnet and the aperture of
slits located after the magnet it was possible
to significantly reduce the $^{12}$C$^{2+}$ contribution at the
expenses of a reduction in the $^6$He$^+$ intensity by a factor of
about 2. The resultant $^{12}$C$^{2+}$ intensity does not affect
the beam preparation before injection into the Paul trap.

\begin{figure}[htb!]
%
\includegraphics[height=22mm]{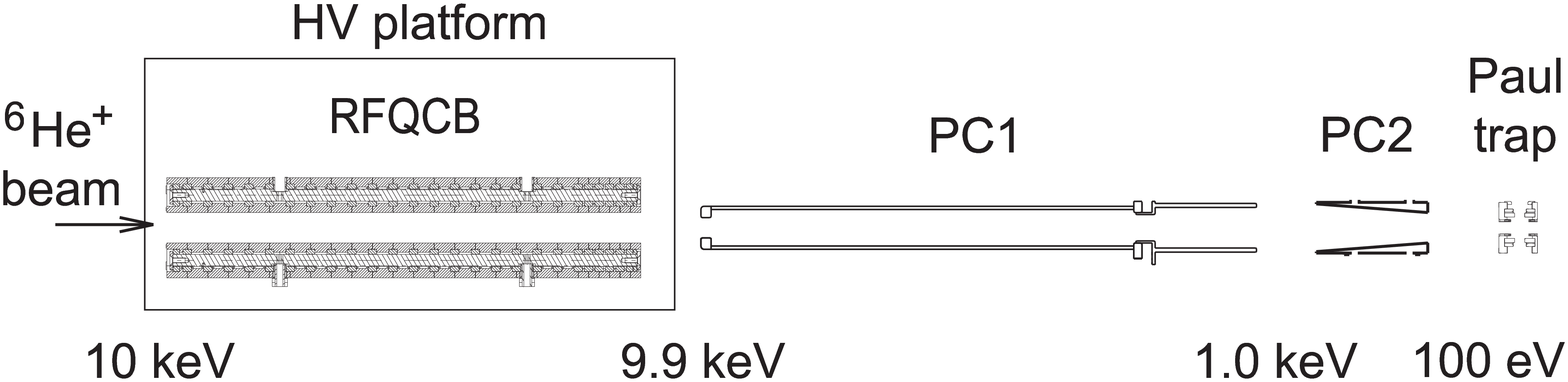}
%
%
\caption{Beam preparation scheme showing the
RFQCB and two pulsed cavities located upstream
from the Paul trap (distances between the
elements are not to scale). The total mean energy of
ions is indicated at different steps.}
\label{fig:beamsetup}
\end{figure}

The setup for the beam preparation (Fig.~\ref{fig:beamsetup})
is comprised of a radio frequency quadrupole cooler and buncher (RFQCB)
followed by two
pulsed electrodes located before the Paul trap.
The beam intensity at the entrance of the RFQCB was
typically 10~nA, including the contribution of the stable
$^{12}$C$^{2+}$ ions. Under optimal conditions, the $^6$He$^+$ beam intensity
was deduced to be $2\times10^8$~s$^{-1}$ 
by implanting a fraction of the beam into a Si detector.

The beam was cooled in the RFQCB using the buffer gas technique \cite{Lunney99},
which is relatively fast and universal,
and well suited for radioactive species.
Since the cooling is only efficient at energies of about 100~eV, the RFQCB
is mounted on a high-voltage platform, operated 100~V below the
voltage of the ECR source platform. In the RFQCB, the ions are confined
radially by an RF field applied to four cylindrical rods. The rods are segmented
in order to generate a longitudinal electrostatic field which drives the ions
toward the exit of the structure. A detailed description of the device
can be found elsewhere \cite{Darius04}.
The cooling of ions as light as $^4$He$^+$
had previously been demonstrated using H$_2$ as buffer gas \cite{Ban04}.
Inside the cooler, the ions are accumulated to produce a bunch for an
efficient
injection in the Paul trap. The bunch is extracted from the RFQCB
by fast switching
the buncher electrodes after thermalization of ions with
the buffer gas.

The ions are then transported
through a first pulsed cavity (PC1) followed
by an electrostatic lens (not shown on Fig.~\ref{fig:beamsetup})
and finally through a
second pulsed cavity (PC2) before their
injection into the Paul trap. Switching voltages applied to PC1 and PC2
reduce the mean ion energies from 9.9~keV to 1~keV and then
from 1~keV to 100~eV respectively \cite{Rodriguez06}
in order to achieve an efficient capture of the
ion bunch by the Paul trap.
 
\begin{figure}[hbt!]
%
\includegraphics[height=68mm]{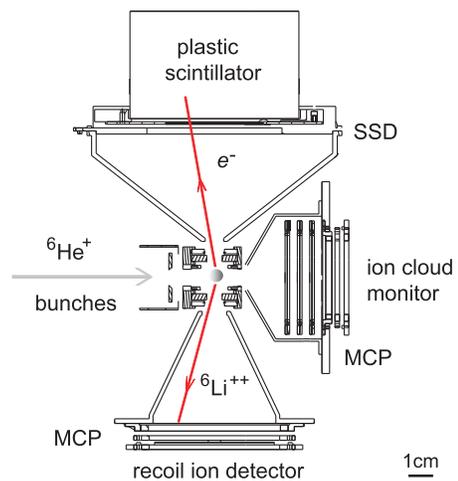}
%
%
\caption{Experimental setup with the transparent Paul trap surrounded by the
beta particle telescope and the two ion detectors.
The scale indicates the size of this table top experiment.}
\label{fig:trap}
\end{figure}

The ion bunches are injected in the Paul trap at a repetition rate
of 10~Hz. The electric field inside the trap is
generated by two pairs of coaxial rings
separated by 10~mm (Fig.~\ref{fig:trap}), and 
is close to a quadrupole field over few mm$^3$ around
the trap center \cite{Rodriguez06}.
The frequency used in the trap was 1.15~MHz for a 130~V peak-to-peak 
voltage amplitude. The RF voltage is applied on the two inner rings
whereas the outer rings are grounded. The RF signal on the trap was
continuously applied during the measuring cycles. The overall performance
of the system has been thoroughly tested with
stable $^4$He$^+$, $^{35}$Cl$^+$ and $^{36,40}$Ar$^+$ ions from the ECR
source \cite{Rodriguez06}
and with $^6$Li$^+$ ions from a surface ionization source \cite{Mery07}.
Considering the duty cycle used for the injection
of the $^6$He$^+$ bunches into the trap, the beam preparation efficiency
was estimated to be $7\times 10^{-5}$, including
the deceleration, cooling, bunching, transmissions through the pulsed
cavities and trapping.

The trap geometry allows the application of suitable voltages on the
rings for the injection and extraction of ions. The absence
of a massive ring electrode also enables the direct
detection of products from decays in the trap.
The trap is surrounded by an electron telescope detector and by two
ion detectors (Fig.~\ref{fig:trap}).
Collimators located in front of the detectors enable the selection
of events originated mainly in the trap.
The number of trapped ions was continuously monitored
by counting the ions remaining in the trap after a fixed storage time,
using the micro-channel plate (MCP) detector located downstream.
This detector is preceded by three grids to reduce the intensity of
the incident ion bunches. 
The measured time-of-flight distribution contains a single peak
corresponding to a mass-to-charge ratio $Q/A = 6$.
The storage time of ions in the trap, deduced from the rate of
coincidence events and accounting for the $\beta$-decay,
was 240~ms for a typical
pressure in the trap chamber of $2\times 10^{-6}$~mbar
due to H$_2$ gas leaking from the RFQCB.
The main effect limiting the storage time of ions inside the trap is the
collision of ions with molecules of the residual gas.

The telescope for beta particles is composed of a $60\times60$~mm$^2$, 300~$\mu$m thick,
double-sided position sensitive silicon strip detector (SSD), with $2\times 60$
strips for horizontal and vertical location. The SSD is followed by a
\O$11\times 7$~cm$^2$
plastic scintillator.
To achieve a better vacuum in the trap chamber,
both detectors are located in an evacuated chamber separated
from the trap chamber by a 1.5~$\mu$m thick mylar foil.
The recoil ion detector uses two MCPs with delay-line readout
providing position sensitivity. The time resolution of the detector is
smaller than 200~ps.
The dependence of the detector efficiency as a function of the ion kinetic
energy, the incidence angle and the position has been studied in
detail and was described elsewhere \cite{Lienard05}. An acceleration voltage is
applied on an electrode located 6~mm in front of the MCP. The ion detection
efficiency reaches 53\% for post-accelerating voltages larger than 4~kV.

The trigger of an event is generated by a signal in the plastic scintillator.
The 120 strips of the SSD are then read-out
sequentially by multiplexing the signals. The trigger also generates the start signal
for the time measurement relative to the MCP
detector located opposite to the trap. Under regular conditions, the average
trigger rate
was 200~s$^{-1}$. Most of these events were singles, for which only the
signal of the energy deposited
in the plastic scintillator was recorded along with the position and
energy information from the 120 strips of the SSD. For coincidence
events, six other parameters were additionally stored:
{\em i)} the time difference between the plastic scintillator
and the MCP signal; {\em ii)} the charge in the MCP;
and {\em iii-vi)} the time differences between the signal from the
MCP and the four outputs of the delay lines (two for each coordinate)
for the position reconstruction. For all events,
the time within the cycle relative to the extraction pulse from the RFQCB
as well as the RF phase in the Paul trap were also recorded
for later study of the ion cloud stability and control
of instrumental effects.

\begin{figure}[htb!]
\centerline{
%
\includegraphics[height=60mm]{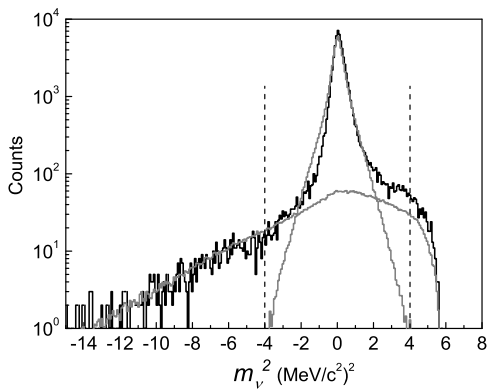}
%
}
\vspace{-5mm}
\caption{Anti-neutrino rest mass spectrum reconstructed from the decay
kinematics (black) and compared with results from MC
simulations (gray).
The main peak corresponds to decay events occurring in the trap. The broad
asymmetric
distribution is due to decays from neutral $^6$He atoms occurring
outside of the trap and to accidental events. See text for details.}
\label{fig:neutrino-rest-mass}
\end{figure}

With the position information of the beta particle and of the recoiling ion, it is
possible to reconstruct the rest mass of the anti-neutrino from the decay
kinematics which
provides a useful control means to identify background sources.
The spectrum built after energy calibration of the plastic scintillator
and selection of beta particles with energies above 1~MeV is shown in
Fig.~\ref{fig:neutrino-rest-mass}.
Two other conditions were imposed to the data: 1)
the signal in the SSD should have a valid conversion in the vertical and horizontal
positions, corresponding to a minimum ionizing particle;
2) the coincidence must be recorded at least 20~ms after the injection of the ion
bunch in the trap, which is needed for the ions to reach the thermal equilibrium. 

The measured spectrum is compared to a Monte-Carlo (MC)
simulation which includes the
geometry of the setup, the size and temperature
of the ion cloud inside the trap, the effect of
the trap RF field on the ions inside the trap and during their recoil,
and the energy resolution of the plastic
scintillator. The simulation does however not include:
the scattering of electrons on the
matter surrounding the trap with the possible generation of secondary background,
the response function of the plastic scintillator, the effect of ionization
during $\beta$-decay \cite{Patyk07} and other recoil order effects beyond
the allowed approximation for the description of the beta
decay process.
For the purpose of this comparison, the value of the
$\beta\bar{\nu}$ angular correlation coefficient 
was assumed to be the one expected from the standard model
for a pure Gamow-Teller transition,
$a_{GT} = -1/3$ \cite{Jackson07}.
The width of the signal peak in Fig.~\ref{fig:neutrino-rest-mass}
is dominated by the phase
space of ions inside the trap \cite{Mery07}. The two main background sources
which have been identified, and which give both rise to
asymmetric distributions in the anti-neutrino rest mass, are 
accidental coincidences and 
decay events of neutral $^6$He atoms produced in the RFQCB which diffuse
into the trap chamber.
Although the simplified MC simulation does not reproduce all details
of the measured spectra, the discrepancies are not crucial to
illustrate two main points: i)
that the signal to background ratio in the region of the peak is
about 100:1 and ii) that the largest fraction of the background is well
identified as associated with decays out of the trap such that the
setup can be improved for future precision measurements.

\begin{figure}[htb!]
\centerline{
%
\includegraphics[height=60mm]{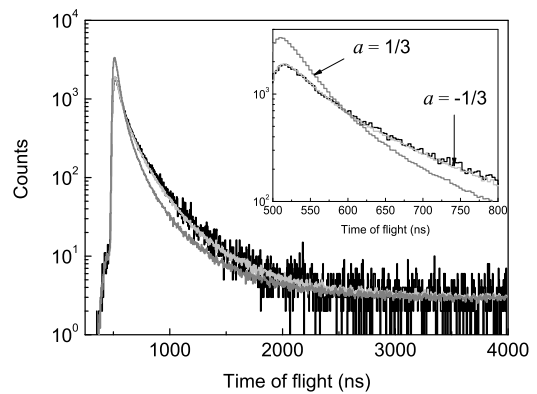}
%
%
}
\vspace{-5mm}
\caption{Comparison between the experimental time-of-flight
spectrum (black) and calculations from MC simulations, assuming
either $a=-1/3$ (light gray) or $a=1/3$ (dark gray). The insert
shows the most sensitive interval to the value of
the $\beta\bar{\nu}$ angular correlation coefficient $a$.}
\label{fig:tof}
\end{figure}

Figure~\ref{fig:tof} shows the time of flight spectrum of
ions relative to the beta particle when selecting events with the condition
$\pm 4~$(MeV/c$^2$)$^2$ on the anti-neutrino rest
mass (Fig.~\ref{fig:neutrino-rest-mass}).
The data correspond to a total measuring time of about 6 days.
Since the overall detection efficiency of
the system is $1.5\times 10^{-3}$
\cite{Mery07}, one concludes that more
than 10$^8$ radioactive ions have been trapped during the measurement.

The data in the time-of-flight spectrum has also been compared to the MC
simulation. The edge of the spectrum, between 460 and 510~ns,
which is weakly sensitive to the dynamics of the decay, served here to accurately
determine the average distance between the ion cloud center and the recoiling
ion detector, and is then not considered in the comparison.
The value of the angular correlation coefficient was here fixed
to either $a_{A} = a_{GT} = -1/3$, corresponding to pure axial
couplings (light gray line in Fig.\ref{fig:tof}), or to $a_{T} = 1/3$
for pure tensor couplings (dark gray line in Fig.\ref{fig:tof}).
Since the number of events in the MC simulation is normalized to the
number of measured events, there are then no free parameters in the
comparison between the data and the MC.
The fairly good agreement between the data and the MC for
$a_{A} = a_{GT}$ proves that the main
features of the setup are
well understood. The most sensitive interval of the spectrum to the
angular correlation coefficient is located between 500 and 800~ns
(see insert in Fig.\ref{fig:tof}), where
the signal-to-accidental ratio is 55. With the data presented here it is
possible to determine
the angular correlation coefficient with a statistical precision
$\Delta a/a = 0.018$, what is a factor of 5 more precise than
the most precise measurement
of this coefficient from a coincidence measurement in $^6$He decay \cite{Vise63},
showing the potential of
this trapping technique. 

The experiment presented above demonstrates 
that abundant quantities of radioactive ions can be efficiently
trapped using a novel transparent Paul trap, and that
their decay products can be recorded with suitable detectors.
The technique has been tested with $^6$He$^+$ ions, which are the lightest
short-lived radioactive ions ever trapped. This required the extension of the buffer gas
cooling technique to the lightest accessible masses \cite{Ban04} and its application
to radioactive
species. The trapping scheme provides significantly larger efficiencies
for noble gas elements
than those obtained so far with MOTs and offers a more
suitable environment than Penning traps for the detection of decay
products. The technique
can be extended to any radioactive ion, 
such as for example $^8$Li$^+$ or $^{19}$Ne$^+$,
opening the possibility for new trap assisted decay experiments.

Since the completion of the experiment presented here, the efficiency
of the beam preparation system
was improved by almost two orders of magnitude \cite{Duval08} enabling 
the possibility for a
high precision measurement of the angular correlation coefficient in
$^6$He decay.

We are grateful to J.~Bregeault,
Ph.~Desrues, B.~Jacquot, Y.~Merrer, Ph.~Vallerand, Ch.~Vandamme and F.~Varenne
for their assistance during the different phases of the project.
We thank J.~Blieck and Y.~Lemi\`ere for their cooperation during
the experiment. This work was supported in part by the R\'egion Basse Normandie,
by the NIPNET RTD network within the
6th FP (contract Nr HPRI-CT-2001-50034) and by
the TRAPSPEC JRA within the I3-EURONS (contract Nr
506065). D.~Rodr\'{i}guez acknowledges support from the E.U. under a
Marie Curie Intra-European Fellowship (contract Nr MEIF-CT-2005-011269).

%

\end{document}